\documentclass{amsart}
\usepackage[dvips]{epsfig}

\usepackage{graphicx}
\usepackage{multirow}
\usepackage{latexsym}
\usepackage{amsmath}
\usepackage{amsthm}
\usepackage{amssymb}
\usepackage{booktabs}
\usepackage{enumitem}
\usepackage{lscape}

\RequirePackage[colorlinks=true,citecolor=blue,urlcolor=blue]{hyperref}
\usepackage{ulem}
\usepackage{xcolor}
\usepackage{caption}
\usepackage{subcaption}
\usepackage{float}
\usepackage{hyperref}
\usepackage{comment}
\usepackage{eurosym}
\newcommand{\vip}{\vskip.2cm}

\newcommand{\E}{\mathbb{E}}
\newcommand{\COMMENTAIRE}[1]{}

\graphicspath{{figures/}}
\title[]{Battery valuation on electricity intraday markets with liquidity costs}
\author{Enzo Cogn\'eville, Thomas Deschatre, \and Xavier Warin}
\address{Thomas Deschatre, EDF Lab Paris-Saclay and FiMe, Laboratoire de Finance des March\'es de l'Energie, 91120 Palaiseau, France}
\email{thomas-t.deschatre@edf.fr}
\address{Enzo Cognéville, EDF Lab Paris-Saclay and FiMe, Laboratoire de Finance des March\'es de l'Energie, 91120 Palaiseau, France}
\email{enzo.cogneville@edf.fr}
\address{Xavier Warin, EDF Lab Paris-Saclay and FiMe, Laboratoire de Finance des March\'es de l'Energie, 91120 Palaiseau, France}
\email{xavier.warin@edf.fr}

\begin{document}
\begin{abstract} 
In this paper, we propose a complete modelling framework to value several batteries in the electricity intraday market at the trading session scale. The model consists of a stochastic model for the 24 mid-prices (one price per delivery hour) combined with a deterministic model for the liquidity costs (representing the cost of going deeper in the order book). A stochastic optimisation framework based on dynamic programming is used to calculate the value of the batteries. We carry out a back test for the years 2021, 2022 and 2023 for the German market and for the French market. We show that it is essential to take liquidity into account, especially when the number of batteries is large: it allows much higher profits and avoids high losses using our liquidity model. The use of our stochastic model for the mid-price also significantly improves the results (compared to a deterministic framework where the mid-price forecast is the spot price). 
\end{abstract}
\maketitle
%\textbf{Mathematics Subject Classification (2020)}: 60G55; 60G57; 62P05; 91G30.

\section{Introduction}

\subsection{Motivation} 
The capacity and production of renewable electricity is increasing every year; in France, for example, the aim is to achieve a 33\% share of renewable energy in electricity consumption in 2030\footnote{\url{https://www.ecologie.gouv.fr}}. However, the producers involved have to deal with  the intermittent nature of these production resources, which are subject to the uncertainties of the weather and to the outages of traditional thermal unit production. In Western Europe, in the spot market, producers submit their hourly production offers for the following day by noon on the day prior. They commit to delivering this production the next day at a predetermined price, known as the spot price, which is set for each hour. This balance is therefore based on a forecast of the previous day's production, and this forecast can change. Producers can use the intraday electricity market to rebalance their positions, which opens at 3 p.m. (CET) the day before delivery and consists of one product for each hour of delivery, which can be bought or sold up to 5 to 30 minutes (depending on the country) before the hour of delivery. For more details on how the intraday market works, we refer the reader to \cite[Section 2]{hirsch2024b}. Another way of coping with this intermittency is to use storage resources such as batteries: the operator can then release energy from storage if there is a shortfall in production and store it otherwise. Batteries can also be used to respond to peaks in demand (or falls in production) by releasing energy from storage, or to very low demand (or excessively high production) by storing energy. To quantify the profitability of the battery, it is necessary to quantify the flexibility of this means of storage on the markets, in this case on the intraday market. Much theoretical work has been devoted to storage valuation~\cite{warin2012, carmona2010, bachouch2022} for example, see also~\cite{machlev2020} for a review, with the most recent being the work of Abi Jaber et al.~\cite{jaber2024} which includes market effects and transaction costs. On the empirical side, Deschatre and Warin~\cite{deschatre2024} as well as Collet et al.~\cite{collet2018} use dynamic programming to value a battery on the electricity intraday market (in~\cite{collet2018}, the battery is managed with a wind generation facility) ; Jiang and Powel \cite{jiang2015} consider the real time market (hour-ahead) in United-States.
%They model a price per maturity and over a time horizon that corresponds to several trading sessions and the optimisation is also carried out using dynamic programming. 

\medskip
In \cite{deschatre2024, collet2018, jiang2015}, liquidity costs are not taken into account. These can have a significant impact, especially when a player uses several batteries simultaneously. In addition, the intraday market can have very high bid-ask spreads, which is a very important indicator of the cost of liquidity, see~\cite{balardy2022, bergault2024}. Neglecting these costs can lead to an overestimation of battery yields and a bad investment. Few papers have focused on modelling liquidity in intraday electricity markets. Favetto~\cite{favetto2019} models the time arrivals of transaction prices with a Hawkes process with time-dependent intensity. Graf von Luckner and Kiesel~\cite{graf2020} consider the arrivals of market orders in the intraday order book and use a bivariate Hawkes process to model these arrivals; the evolution of the parameters of this model is studied in~\cite{blasberg2019}. In~\cite{kramer2021}, Kramer and Kiesel extend the model of Graf von Luckner and Kiesel~\cite{graf2020} by including exogenous factors such as forecasting errors in renewable energy production. They also consider limit and cancel orders independently. While all of these models provide insights into the behaviour of liquidity in the intraday electricity market, they do not allow for the simulation of trading strategies or backtesting, which requires the simulation of the entire order book (or at least the mid-price with liquidity costs). Recently, Bergault and Cogn\'eville~\cite{bergault2024} have proposed an order book modelling framework for illiquid markets, which is applied to French and German intraday electricity markets and can then be used to simulate trading strategies or asset pricing under liquidity costs. Unfortunately, the last model of \cite{bergault2024} (but also all the previous cited models on liquidity) only considers a model for a given maturity and does not model the dependency between different order books, which is essential for pricing a storage asset. Glas et al.~\cite{glas2020} study the limit order book and model the cost of executing a market order as a linear function of volume with time-dependent parameters. They use their model to optimise the revenue of a portfolio of conventional and wind generation sold on the intraday electricity market. Kath and Ziel~\cite{kath2020} follow the same approach using Generalized Additive Models~\cite{wood2017} to model the execution cost of a market order and solve the optimal execution problem. 

\subsection{Contribution} Our approach follows that of Glas et al.~\cite{glas2020} or Kath and Ziel~\cite{kath2020} and then differs from that of Bergault and Cognéville~\cite{bergault2024}. For our case, modelling the entire order book across all 24 maturities in the EPEX market would result in an excessively high-dimensional framework, with 6 processes required for each maturity. Instead, we propose a model for mid-prices and a model for liquidity costs. This approach was first proposed for equity markets by Cetin et al. in \cite{cetin2006} and Blais and Protter in \cite{blais2010}. They consider a share price per unit of volume that a trader pays/receives when he buys/sells $x$ units of share in the market at time $t$ (where $x>0$ corresponds to a buy and $x<0$ to a sell). The total price paid by the trader is then $xS(t,x)$ (which is negative if he receives money). Cetin et al.~\cite{cetin2006} consider the multiplicative model $S(t,x)=S(t,0)e^{\alpha x}$, and $x \to e^{\alpha x}$ is a supply curve corresponding to the cost of liquidity: the more you buy, the deeper you have to go in the order book and the more you pay per unit of volume. $S(t,0)$ follows a Black-Scholes dynamic. Blais and Protter~\cite{blais2010} consider an additive model 
\begin{equation}
    S(t,x) = S(t,0) + M(t)x \label{eq:prottermodelliquid}
\end{equation}
and for illiquids an asymmetric model 
\begin{equation}
    S(t,x) = (b_-(t)+ M_-(t)x){\bf 1}_{x < 0} + (b_+(t)+ M_+(t)x){\bf 1}_{x \geq 0}. \label{eq:prottermodel}
\end{equation} The parameters of the liquidity curve in \cite{blais2010} are estimated using the limit order book.

\medskip
In this paper, we propose a model of both price and liquidity for intraday electricity prices that allows for the pricing of assets such as storage, whose value depends on the dynamics of prices for different maturities. We first perform an empirical analysis using limit order book data from EPEX for each maturity and analyse the liquidity curve (defined around the mid-price, as in~\cite{malo2012}). We fit a model close to \eqref{eq:prottermodel} for the liquidity curve, which is more appropriate than \eqref{eq:prottermodelliquid} and shows evidence of market illiquidity. Our model takes into account some peculiarities of the intraday electricity markets, such as the increase in liquidity as maturity approaches (Samuelson effect)~\cite{balardy2022}. These results are given in Section~\ref{sec:modelling_liquidity}. For the dynamics of mid-prices, we consider the multivariate model of Deschatre and Warin~\cite{deschatre2024}, which has already been used for transaction prices and allows to model several stylised facts observed for intraday electricity prices: volatility increasing as maturity approaches and correlation decreasing with the distance between two maturities, see Section~\ref{sec:modelling_price}. We provide orders of magnitude for the different parameters of the models (liquidity model and mid-price model), which may be useful for researchers and practitioners. These parameters can be used, for example, to study optimal trading strategies or equilibrium pricing~\cite{aid2016, feron2020, aid2022}, which require volatility parameters for the price and transaction costs. An alternative to the Deschatre and Warin model~\cite{deschatre2024} for mid-price simulation would be the model of Hirsch and Ziel~\cite{hirsch2024b} used for probabilistic forecasting, but it requires a lot of information such as wind power forecast as input.

\medskip
To assess the quality of the liquidity model, we consider the valuation of a storage, a battery here, in Section~\ref{sec:optim}. We perform a backtest on real data for the French and German electricity intraday markets and compare the revenue induced by optimising the storage asset using our liquidity model and without liquidity model. Applying the optimal controls obtained with a dynamic programming algorithm on real data, we show that revenues are higher using a control that takes liquidity into account.
Neglecting those costs can lead to possible high losses when the number of battery is high. It is therefore essential to include a liquidity model in the optimisation process. We also show that the use of the stochastic model improves the battery's value.

\subsection{Dataset} The dataset used consists of all orders for each trading session during 2021, 2022 and 2023 on the French and German intraday electricity markets, provided by the market platform EPEX. For each delivery date, we have access to orders from 24 trading sessions covering the 24 delivery hours (or maturities, or products) of the day (we do not consider products, which also exist, with half-hourly or quarter-hourly delivery). These sessions start at 3 p.m. (CET) the day before delivery and end 5 minutes before the delivery hour (the trading duration is then different for each maturity). For a given maturity, we have access to all limit and market orders sent to the market. Each order is marked with the time of its creation, to the millisecond, a price, a volume, a side (buy or sell). If the order is cancelled at any time during the trading session, the cancellation date is recorded, which allows us to reconstruct cancelled orders. Some orders may have trading restrictions that are taken into account: Immediate or Cancel (any part of the order not filled immediately is cancelled), Fill or Kill (executed immediately at a specified price or cancelled if not filled in full), All or None (the order is filled in full or cancelled). In addition, some orders can be hibernated, i.e. deactivated and then reactivated; and some orders are block orders and link different markets: a block order is an order placed simultaneously on several products, the volume of which can only be executed on all products at once; block orders can only be executed between themselves. We do not take block orders into account. Cross-border trading is possible through the Cross-Border Intraday Initiative (XBID project), which allows order books to be aggregated across interconnected countries in Europe as long as the interconnections are not saturated. One hour before delivery, this XBID mechanism closes and cross-border trading is no longer possible, which has a major impact on liquidity and trading: we do not consider this last hour in this study. The tick size is 0.01\euro/MWh. The reader can refer to~\cite[Appendix A]{grindel2022} for a more detailed description of the data.

\medskip
The limit order book is reconstructed from these orders using methods close to the one described in~\cite[Section 3]{grindel2022}: it consists of a snapshot at each point in time of the different orders that can be bought or sold in the market, with their associated price and volume. Note that the limit order book reconstruction may not be exact for some trading sessions as some information is missing, especially on hibernated orders, see also~\cite{grindel2022, finhold2023} which highlights issues with the reconstruction. For France and Germany respectively, we only keep the 20 and 60 best offers available on the bid and ask side at any given moment during the trading session, which is sufficient for this work (also, for the majority of traded products, the dataset often contains less than 20 bid and ask offers for the French market and less than 60 offers for the German market). 

\section{Modelling framework}

Let $0 < T_1<T_2<\cdots<T_M = T$ be the different maturities, with $M \in \mathbb{N} \setminus \{0\}$. We denote by $S_{m,t}(x)$ the price per unit of volume that a trader pays / receives when he buys / sells at date $t$, $x$ MWh of electricity (where $x>0$ corresponds to a purchase and $x<0$ to a sale) delivered at maturity $T_m$, the total price paid by the trader being $xS_{m,t}(x)$. This price is decomposed into the sum of two components: 
\begin{itemize}
    \item[-] the mid-price, denoted by $f_{m,t}$, which is modelled by a stochastic process in Section~\ref{sec:modelling_price},
    \item[-] and the cost of liquidity, $L_m(t,x)$, modelled by a deterministic curve in Section~\ref{sec:modelling_liquidity}.
\end{itemize}

\subsection{Liquidity costs modelling}
\label{sec:modelling_liquidity}

\begin{figure}[H]
     \centering
     \includegraphics[width=0.7\linewidth]{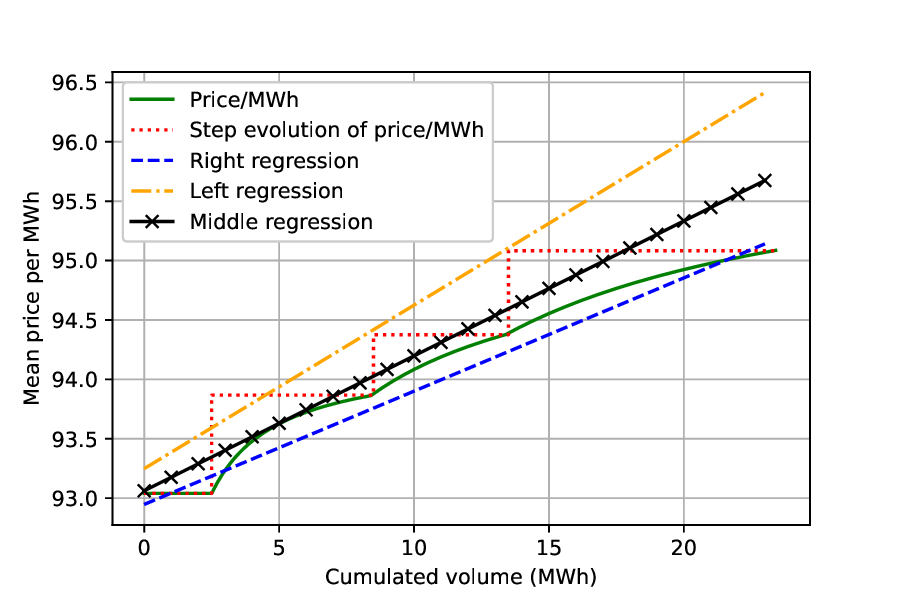}
     \caption{Different ways of looking at the liquidity cost per unit volume: the green plain curve corresponds to the real price per unit volume defined in~\eqref{eq:liq_cost_emp}, the red dashed curve corresponds to the stepwise function defined in~\eqref{eq:liq_cost_emp_stepwise}, the black line with x markers is the linear model~\eqref{eq:liq_cost_model} with parameters estimated in the regression~\eqref{eq:min_param_time}, the orange dashed (resp. blue dashed curve) is the same with parameters estimated in regression~\eqref{eq:min_param_time} using the points $(\bar{V}_{m,i}(t), \bar{p}_m(t,\bar{V}_{m,i+1}(t)))$ (resp. $(\bar{V}_{m,i}(t), \bar{p}_m(t,\bar{V}_{m,i}(t)))$) instead of the points $(\frac{\bar{V}_{m,i}(t) + \bar{V}_{m,i+1}(t)}{2}, \bar{p}_m(t,\frac{\bar{V}_{m,i}(t) + \bar{V}_{m,i+1}(t)}{2}))$.}
     \label{fig:shift}
\end{figure}

To capture liquidity costs in \euro/MWh, we use a linear jump model inspired by the framework developed by Blais and Protter \cite{blais2010} that generalised the linear model used by~\cite{glas2020} for intraday electricity prices. The model is expressed as a function of time $t$ and traded volume $x$:
\begin{equation}
\label{eq:liq_cost}
L_m(t,x) = [A_{m,-}(t)x - B_{m,-}(t)]{\bf 1}_{\{x<0\}} + [A_{m,+}(t)x + B_{m,+}(t)]{\bf 1}_{\{0  < x\}},\,t\leq T_m
\end{equation}
where
\begin{itemize}
    \item[-] $A_{m,-}(t)$ and $A_{m,+}(t)$ are linear functions of the time to maturity $T_m-t$, 
    \begin{equation} \label{eq:A}
        A_{m,\pm}(t) = \alpha_{A_{m,\pm}} (T_m-t) + \beta_{A_{m,\pm}},
    \end{equation}
    \item[-] and $B_{m,-}(t)$ and $B_{m,+}(t)$ are exponential functions, 
    \begin{equation} \label{eq:B}
    B_{m,\pm}(t) = e^{\alpha_{B_{m,\pm}} (T_m-t) + \beta_{B_{m,\pm}}}.
    \end{equation}
\end{itemize}   
To support this model, we first define the empirical counterpart of $L_m(t,x)$. 
Let $V_{m,i}(t)$ with  $V_{m,i}(t) <0$ if $i \le -1$, $V_{m,i}(t) >0$ if $ i \ge 1$, $V_{m,0} =0$, and $P_{m,i}(t)$ be the different volumes and prices offered on the ask (resp. bid) side at time $t < T_m$ for maturity $T_m$, with $P_{m,i}(t) < P_{m,j}(t)$ if $j > i$. We also denote for $i \ge 1$ the cumulative volumes by $\bar{V}_{m,i}(t)$ and $\bar{V}_{m,-i}(t)$, i.e. $\bar{V}_{m,i}(t) = \sum_{k=1}^i V_{m,k}(t)$, $\bar{V}_{m,-i}(t) = \sum_{k=1}^i V_{m,-k}(t)$ and also $\bar{V}_{m,0}(t) = 0$. The empirical counterpart of $L_m(t,x)$ in the Equation~\eqref{eq:liq_cost} is given by 
\begin{equation} \label{eq:liq_cost_emp}
\begin{split}
&p_m(t,x) = \\
&\begin{cases} 
P_{m,1}(t) - \frac{P_{m,-1}(t) + P_{m,1}(t)}{2} & \text{if }  0 < x \leq \bar{V}_{m,1}(t) \\[10pt]
\frac{\sum_{i=1}^{j} P_{m,i}(t) V_{m,i}(t) + (x-\bar{V}_{m,j}(t))P_{m,j+1}(t)}{x}  - \frac{P_{m,-1}(t) + P_{m,1}(t)}{2}& \text{if }  \bar{V}_{m,j}(t) < x \leq \bar{V}_{m,j+1}(t),\,j \geq 1 \\[10pt]
P_{m,-1}(t) - \frac{P_{m,-1}(t) + P_{m,1}(t)}{2} & \text{if }  \bar{V}_{m,-1}(t) \leq x < 0\\[10pt]
\frac{\sum_{i=1}^{j} P_{m,-i}(t) V_{m,-i}(t) + (x-\bar{V}_{m,-j}(t))P_{m,-j-1}(t)}{x}  - \frac{P_{m,-1}(t) + P_{m,1}(t)}{2}& \text{if }  \bar{V}_{m,-j-1}(t) \leq x < \bar{V}_{m,-j},\,j \geq 1 \\[10pt]
\end{cases}
\end{split}
\end{equation}
%or written differently 
%\begin{equation} \label{eq:liq_cost_emp}
%p_m(t,x) = \begin{cases} 
%P_{m,1}(t)  - \frac{P_{m,-1}(t) + P_{m,1}(t)}{2} & \text{if }  0 < x \leq \bar{V}_1(t) \\
%P_{j+1}(t) - \frac{\sum_{i=1}^j (P_{j+1}(t) - P_i(t))V_i(t)}{x}  - \frac{P_{m,-1}(t) + P_{m,1}(t)}{2}& \text{if }  \bar{V}_{j}(t) < x \leq \bar{V}_{j+1}(t),\,j \geq 1 \\
%P_{m,-1}(t) - \frac{P_{m,-1}(t) + P_{m,1}(t)}{2} & \text{if }  \bar{V}_{-1}(t) \leq x < 0\\
%P_{-j-1}(t) - \frac{\sum_{i=1}^j (P_{-j-1}(t) - P_{-i}(t))V_{-i}(t)}{x} - \frac{P_{m,-1}(t) + P_{m,1}(t)}{2}& \text{if }  \bar{V}_{-j-1}(t) \leq x < \bar{V}_{-j}(t),\,j \geq 1 \\
%\end{cases}
%\end{equation}
which is stepwise concave with rupture points at each $\bar{V}_{m,\pm i}(t)$ (see the green plain curve in Figure~\ref{fig:shift} for an example) ; the second term in each case is the mid-price.

\begin{figure}[H]
    \centering
    \begin{tabular}{cc}
     \includegraphics[width=0.45\linewidth]{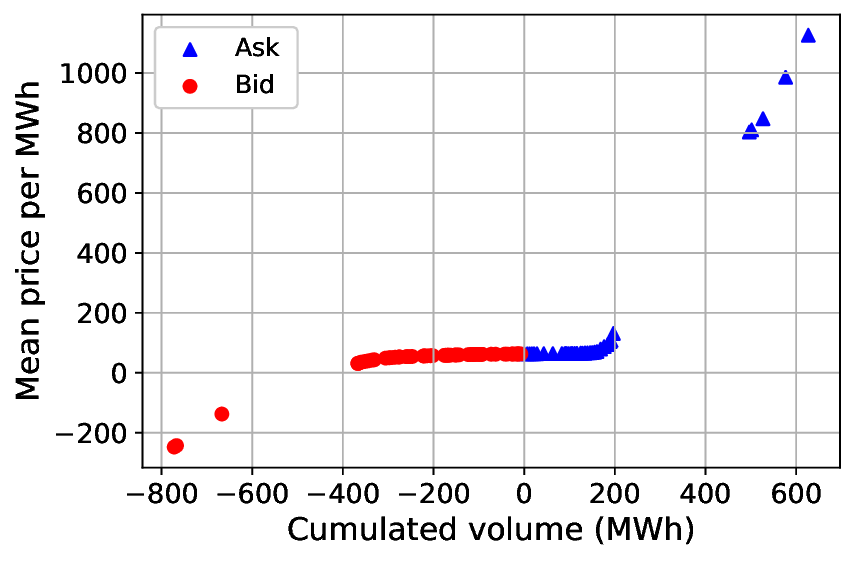}    &  \includegraphics[width=0.45\linewidth]{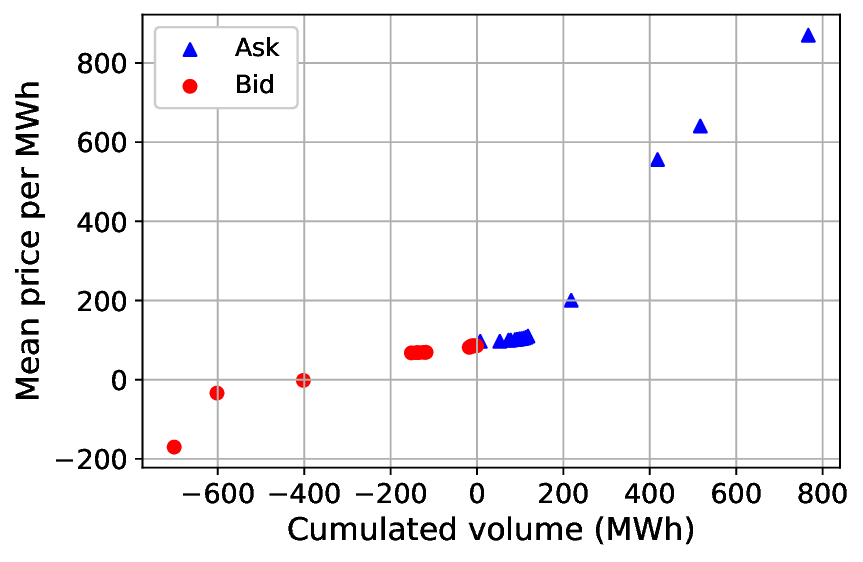}\\
         (A) Germany, 2021-02-28 & (B) France, 2021-01-07\\
      \includegraphics[width=0.45\linewidth]{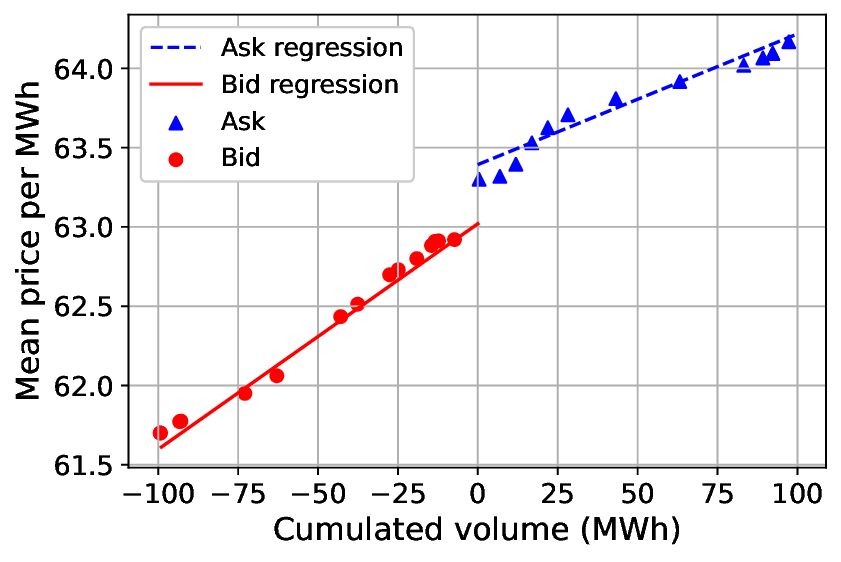} &       \includegraphics[width=0.45\linewidth]{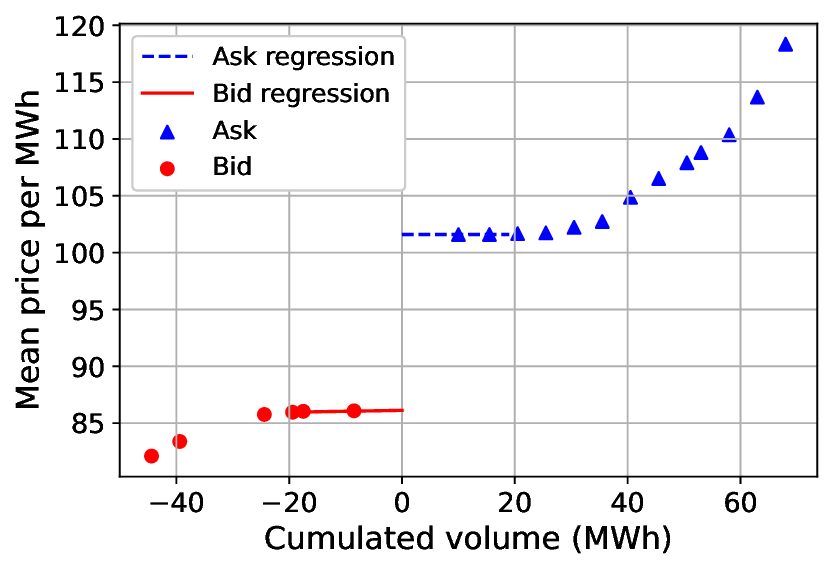} \\
      (C) Germany, 2021-02-28, zoom & (D) France, 2021-01-07, zoom
    \end{tabular}    
    \caption{Example of empirical liquidity curves $p_m(t,\bar{V}_{m,i}(t))$ (points), $i\neq0$, with $p_m$ defined in~\eqref{eq:liq_cost_emp} on German and French market for maturity 18h, $t=3$ hours before maturity, with regression model~\eqref{eq:liq_cost_model}-\eqref{eq:min_param_time} (straight lines) and parameters estimated for $|\bar{V}_{m,i}| \leq 100$ MWh on German market and $|\bar{V}_{m,i}| \leq 20$ MWh on French market.}
    \label{fig:liquidity_curves}
\end{figure}

\medskip
In Figure~\ref{fig:liquidity_curves} we first plot the different values of $p_m(t, \bar{V}_{m,\pm i}(t))$ for $i \geq 1$. We employ the linear jump model~\eqref{eq:liq_cost} to analyze trading volumes within the ranges of $[-20, 20]$ MWh for the French market and $[-100, 100]$ MWh for the German market. These volume intervals are derived empirically for the linear approximation to be valid, as illustrated in Figure \ref{fig:liquidity_curves}. Consequently, our battery valuation in Section~\ref{sec:optim} will be confined to these specified trading volumes. A more precise optimal value window could be determined by examining the regression's residual error value as the volume increases. If one wants to model prices deeper in the order book, it is of course possible to use higher order polynomials or non-parametric functions as in~\cite{kath2020}. For these reasons, we focused our work on these smaller volumes and proceeded to estimate the relevant parameters for the linear jump model. We then display the function
\begin{equation}
\label{eq:liq_cost_model}
\hat{L}_m(t,x) = [\hat{A}_{m,-}(t)x - \hat{B}_{m,-}(t)]{\bf 1}_{\{x<0\}} + [\hat{A}_{m,+}(t)x + \hat{B}_{m,+}(t)]{\bf 1}_{\{0<x\}},\,t\leq T_m
\end{equation}
with $\hat{A}_{m,\pm}(t)$ and $\hat{B}_{m,\pm}(t)$ solutions of 
\begin{equation} \label{eq:min_param_time}
\underset{    \begin{array}{c}
A_{m,\pm}(t) \geq 0, \\
B_{m,\pm}(t)\geq 0
\end{array}
}{\text{argmin}} 
 \sum_{i} \left(L_m\left(t,\frac{\bar{V}_{m,i-1}(t) + \bar{V}_{m,i}(t)}{2}\right) - \bar{p}_m\left(t,\frac{\bar{V}_{m,i-1}(t) + \bar{V}_{m,i}(t)}{2}\right)\right)^2 {\bf 1}_{|\bar{V}_{m,i}(t)| \leq K}
\end{equation}
where $K$ is $20$MWh for the French market and $100$MWh for the German market and
\begin{equation} \label{eq:liq_cost_emp_stepwise}
\bar{p}_m(t,x) = \begin{cases} 
P_{m,1}(t)  - \frac{P_{m,-1}(t) + P_{m,1}(t)}{2} & \text{if }  0 < x \leq \bar{V}_{m,1}(t) \\[10pt]
\frac{\sum_{i=1}^{j+1} P_{m,i}(t) V_{m,i}(t)}{\bar{V}_{m,j+1}}  - \frac{P_{m,-1}(t) + P_{m,1}(t)}{2}& \text{if }  \bar{V}_{m,j}(t) < x \leq \bar{V}_{m,j+1}(t),\,j \geq 1 \\[10pt]
P_{m,-1}(t)  - \frac{P_{m,-1}(t) + P_{m,1}(t)}{2} & \text{if }  \bar{V}_{m,-1}(t) \leq x < 0\\[10pt]
\frac{\sum_{i=1}^{j+1} P_{m,-i}(t) V_{m,-i}(t)}{\bar{V}_{m,-j-1}}  - \frac{P_{m,-1}(t) + P_{m,1}(t)}{2}& \text{if }  \bar{V}_{m,-j-1}(t) \leq x < \bar{V}_{m,-j},\,j \geq 1 \\[10pt]
\end{cases}
\end{equation}
is the piecewise constant function taking the value $p_m(t,\bar{V}_{m,i+1}(t))$ for $x \in \left]\bar{V}_{m,i}(t), \bar{V}_{m,i+1}(t)\right]$ and $p_m(t,\bar{V}_{m,-i-1}(t))$ for $x \in \left[\bar{V}_{m,-i-1}(t), \bar{V}_{m,-i}(t)\right[$ and is then an upper bound for $p_{m}(t,x)$ for $x > 0$ and a lower bound for $x < 0$ (see dotted curve in Figure~\ref{fig:shift} for an example). The choice of using the points $(\frac{\bar{V}_{m,i-1}(t) + \bar{V}_{m,i}(t)}{2}, \bar{p}_m(t, \frac{\bar{V}_{m,i-1}(t) + \bar{V}_{m,i}(t)}{2})$ for the regression is mostly empirical. Dealing with the points $(\bar{V}_{m,i}(t), \bar{p}_m(t, \bar{V}_{m,i}(t))$ for the regression can lead to an underestimation of liquidity costs and can affect forecasts and trading decisions, especially when bids are scarce and gaps between bids are large (right regression in Figure~\ref{fig:shift} for an example). On the other hand, using $(\bar{V}_{m,i}(t), \bar{p}_m(t, \bar{V}_{m,i+1}(t))$ can overestimate liquidity costs (left regression in Figure~\ref{fig:shift}). While it's possible to use real liquidity cost from the true curve with $p_m$ instead of $\bar{p}_m$, this approach removes the flexibility to choose between underestimating or overestimating liquidity costs: the further right you go in the step curve, the greater the underestimation, while moving to the left leads to overestimation. The slopes of these order book curves fluctuate randomly, and rare but significant shifts in the shape of these curves have been observed during periods of heightened market stress. The theory proposed by Blais and Protter \cite{blais2010} would suggest that supply curves tend towards a linear slope towards the end of a trading session due to increasing liquidity, which is not entirely consistent with our observations. This discrepancy is likely due to the persistence of illiquidity towards the end of the session and the fact that we consider deeper layers of the limit order book than Blais and Protter. Furthermore, in practical optimisation applications, a linear model tends to produce suboptimal trading decisions, because the spread is neglected. One solution to this problem is to constrain the algorithm to avoid trading at prices within the spread. Another solution, which was chosen here, is to create a liquidity cost model that separates the buy side from the sell side while maintaining a common spread value. This allows appropriate limits to be set for buying and selling.

\begin{figure}[H]
    \centering   
    \begin{tabular}{cc}
      \includegraphics[width=0.45\linewidth]{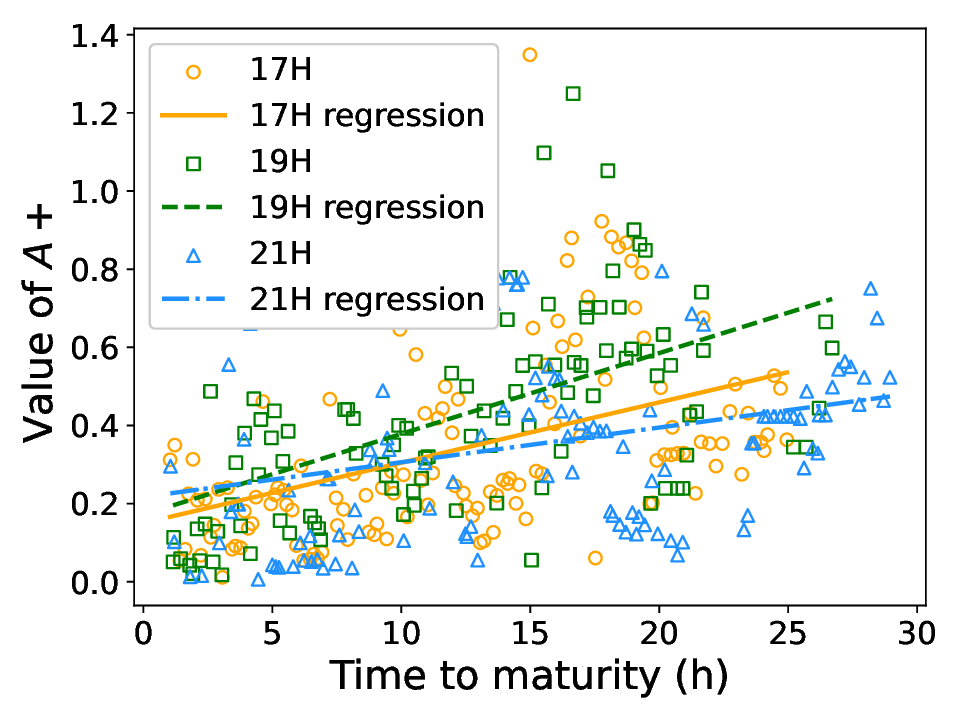} &       \includegraphics[width=0.45\linewidth]{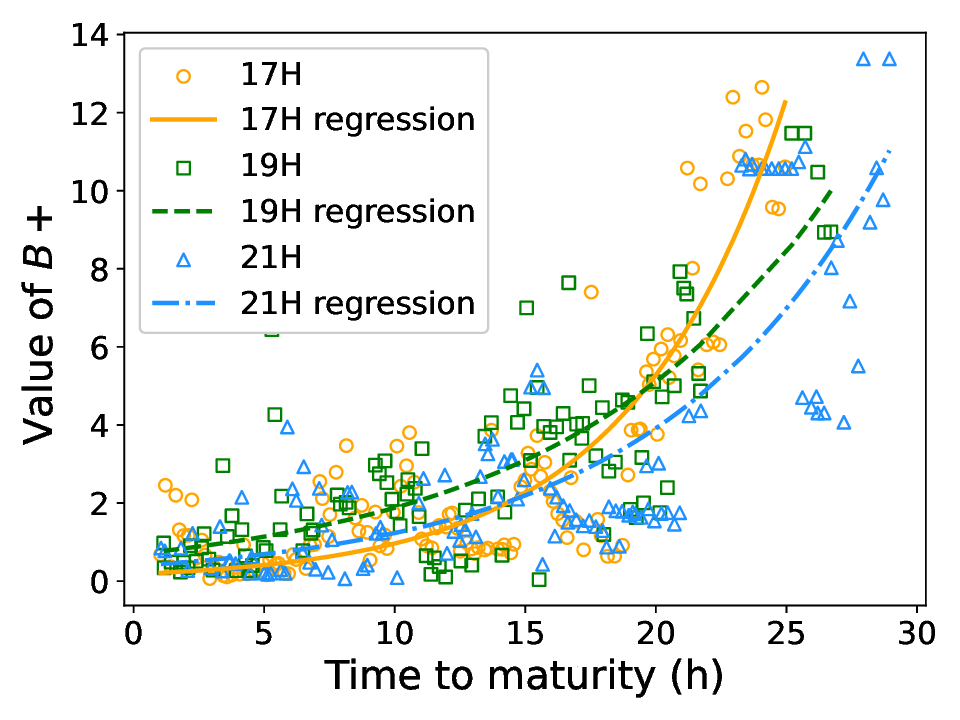} \\
      (A) $A_{m,+}$ & (B) $B_{m,+}$  \\
      \includegraphics[width=0.45\linewidth]{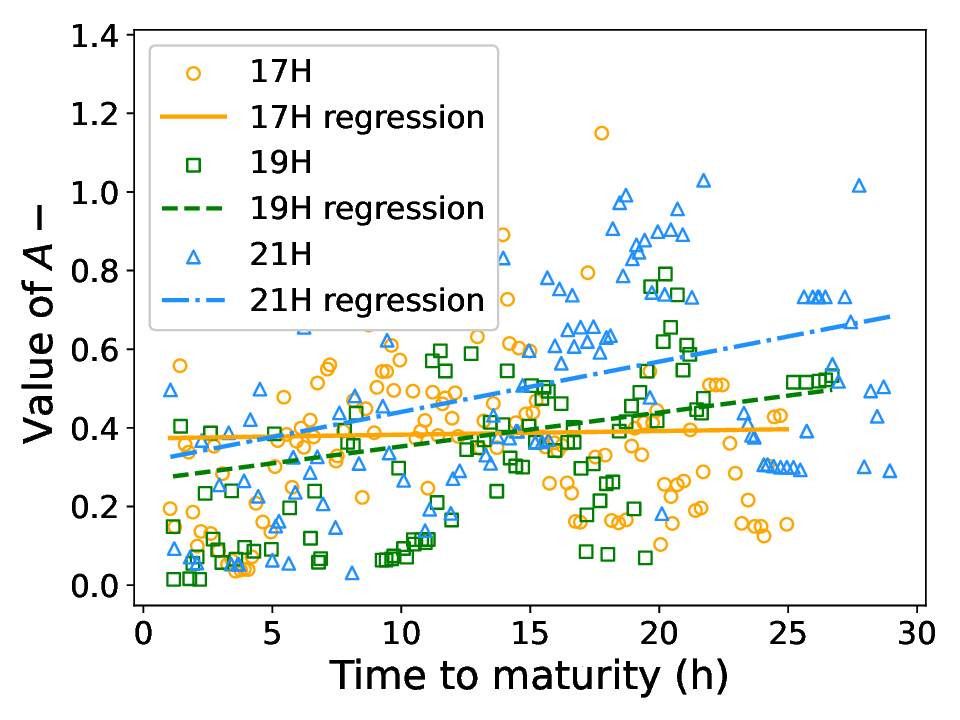} &       \includegraphics[width=0.45\linewidth]{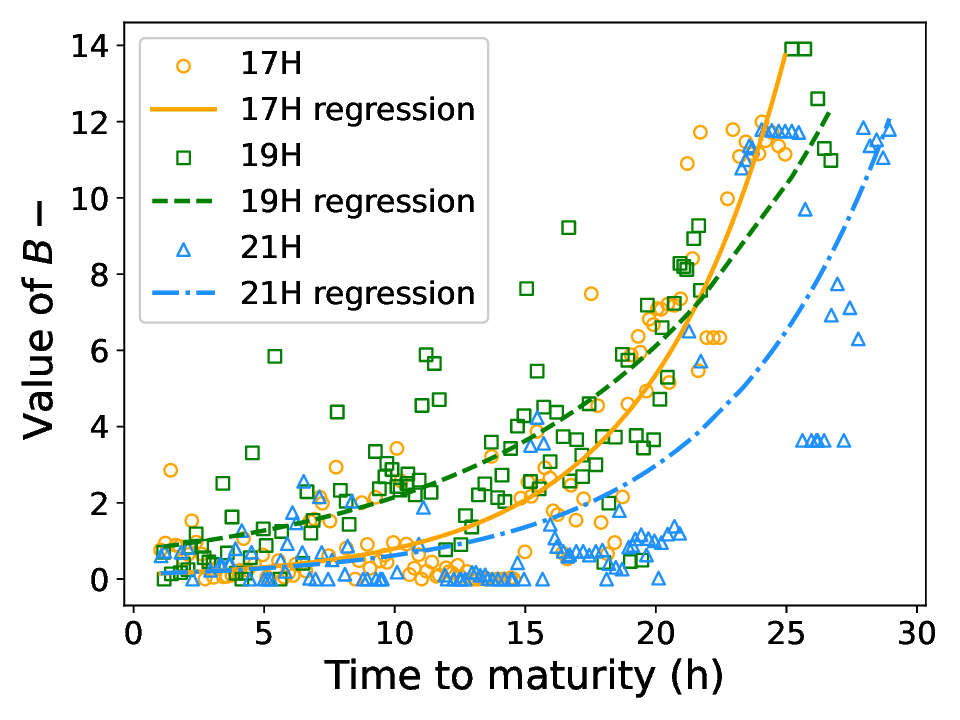} \\
      (C) $A_{m,-}$ & (D) $B_{m,-}$
    \end{tabular}
     \caption{Temporal trends of parameters $\hat{A}_{m,\pm}$ and $\hat{B}_{m,\pm}$ estimated with~\eqref{eq:min_param_time} at each time where there is a change in the order book and when there is at least 5 orders on each side for different maturities $T_m$ and for the French market in January 2021. Each point represents the average parameter value of a given maturity over 10-minute intervals. We fit the parametric models~\eqref{eq:A} and~\eqref{eq:B} to the different points (lines).}
    \label{fig:params}
    
\end{figure}

\medskip
In Figure~\ref{fig:params}, we estimate the quantities $\hat{A}_{m,\pm}(t)$ and $\hat{B}_{m,\pm}(t)$ using Equation \eqref{eq:min_param_time}, on French market data from January 2021. For each time $t$ when a new trade occurred in the market, we computed a set of estimated parameters. To smooth the results, we applied a rolling time window of 10 minutes, taking the average of the estimated parameters within each window. This process produced the plotted results. Subsequently, we performed a regression on all the mean points to capture their dependence on $t$. Note that the bid-ask spread for the maturity $T_m$ at time $t$ is given by $\underset{x \to 0^+}{\lim} L_m(t,x) - \underset{x \to 0^-}{\lim} L_m(t,x) = B_{m,+}(t) + B_{m,-}(t)$ and decreases exponentially with time to maturity, consistent with the findings of Balardy \cite{balardy2022}. The slopes of $A_{m,\pm}(t)$ also decrease as time to maturity decreases, supporting an increase in liquidity.

%\begin{figure}[H]
%     \centering
%     \includegraphics[width=0.5\linewidth]{figures/shift_exemple.eps}
%     \caption{Comparison of shifted and non-shifted regression for positive cumulated volume (buy). Non-shifted estimation tends to underestimate price impact. }
%     \label{fig:shift}
%\end{figure}

%
%\begin{figure}[H]
%    \centering
%    \includegraphics[width=\linewidth]{figures/all_regs_france_2021.eps}

%    \caption{\pg{Sur la plupart des figures, mais notamment celle-ci, les labels et les légendes sont difficilement lisibles / impossibles à lire.}Temporal trends of parameters $\hat{A}_{m,\pm}$ and $\hat{B}_{m,\pm}$ estimated with~\eqref{eq:min_param_time} at each time where there is a change in the order book and when there is at least 5 orders on each side for different maturities $T_m$ and for the French market in January 2021. Each point represents the average parameter value of a given maturity over 10-minute intervals. We fit the parametric models~\eqref{eq:A} and~\eqref{eq:B} to the different points (lines).}
%    \label{fig:params}
%\end{figure}

To estimate the model in the battery valuation framework of Section~\ref{sec:optim}, we consider the different empirical curves $\bar{p}^d_m(t,x)$ defined in~\eqref{eq:liq_cost_emp_stepwise} for each trading session day $d=1,\ldots,D$, where $D$ represents the total number of distinct trading sessions corresponding to the same hour of delivery. Let $(\tau^d_{m,i})_{i \geq 1}$ be the times at which there is a change in the order book fo session $d$ and maturity $m$ and $\bar{V}_{m,i}^d$ be the corresponding cumulative volumes. The various parameters $\alpha_{A_{\pm}, m}$, $\beta_{A_{\pm}, m}$, $\alpha_{B_{\pm}, m}$ and $\beta_{B_{\pm}, m}$ are obtained as the solution of
\begin{equation} 
\begin{split}\label{eq:min_params}
\underset{    \begin{array}{c}
\alpha_{A_{m,\pm}} \geq 0,\,\beta_{A_{m,\pm}} \geq 0, \\
\alpha_{B_{m,\pm}} \geq 0,\,\beta_{B_{m,\pm}}
    \end{array}
}{\text{argmin}} \sum_{\begin{array}{c} d=1,\ldots,D,\\ \tau^d_{m,k},\ i \end{array}} &{\bf 1}_{|\bar{V}^d_{m,i}(\tau^d_{m,k})| \leq K} \left(L_m\left(\tau^d_{m,k},\frac{\bar{V}^d_{m,i-1}(\tau^d_{m,k}) + \bar{V}^d_{m,i}(\tau^d_{m,k})}{2}\right)\right.\\
&\left.-  \bar{p}^d_m\left(\tau^d_{m,k},\frac{\bar{V}^d_{m,i-1}(\tau^d_{m,k}) + \bar{V}^d_{m,i}(\tau^d_{m,k})}{2}\right)\right)^2.
\end{split}
\end{equation}
To match the constraints of financial modelling and observations, we set parameters such that $\alpha_{A_{m\pm}},\, \beta_{A_{m,\pm}},\, \alpha_{B_{m,\pm}} \geq 0$ to ensure positive and negative price effects that diminish as the time to maturity decreases.  In our battery valuation framework, we wanted to optimise trading decisions by making them one or two hours before delivery. To achieve this, we set the time to maturity to $T_m-t = 1$ hour for decisions made one hour before delivery and to $T_m-t=2$ hours for those made two hours before maturity. To improve accuracy, we refined our dataset by focusing on specific time windows: for the two-hour scenario, we included observations from $T_m-t =1.5$ to $T_m-t=2.5$ hours before delivery, while for the one-hour scenario we used data from $T_m-t=1$ to $T_m-t=1.5$ hours. Notably, data between $T_m-t=0.5$ and $T_m-t=1$ hours were excluded, as the XBID electricity market is closed during the last hour before delivery.

\medskip
For practitioners wishing to replicate our price impact simulations, sample parameter sets are provided in Table ~\ref{tab:params_liq_germany} and Table~\ref{tab:params_liq_france}. Note that these parameters have been estimated by regressing the different values of $\hat{A}_{m,\pm}(t)$ and $\hat{B}_{m,\pm}(t)$ obtained from regression~\eqref{eq:min_param_time} averaged over 10 minutes windows against the theoretical model \eqref{eq:A}-\eqref{eq:B} (using regression \eqref{eq:min_params} over the whole trading session would result in a model that poorly represents the early stages of the trading session, as the majority of data is concentrated toward the end of the session when liquidity is highest). The parameters presented in the tables align well with market observations: there is a noticeable increase in liquidity during the beginning of 2022 (war in Ukraine and nuclear plants shutdowns in France at the end of 2021), followed by a slight decrease in 2023, highlighted by the evolution of the parameters of the spreads, $\alpha_{B_{m,\pm}}$ and $\beta_{B_{m,\pm}}$, and of the parameters of the slopes, $\alpha_{A_{m,\pm}}$ and $\beta_{A_{m,\pm}}$.

\begin{table}[H]
\centering
        \begin{tabular}{rrrrrrrrr}
\toprule
Year & $\alpha_{A_{m,+}} T_m$ & $\beta_{A_{m,+}}$ & $\alpha_{B_{m,+}}T_m$ & $\beta_{B_{m,+}}$ & $\alpha_{A_{m,-}}T_m$ & $\beta_{A_{m,-}}$ & $\alpha_{B_{m,-}}T_m$ & $\beta_{B_{m,-}}$ \\
\midrule
2021 & 0.1751 & 0.0122 & 2.6968 & -1.8208 & 0.0859 & 0.0240 & 3.6898 & -2.2508 \\
2022 & 1.1047 & 0.0444 & 45.1306 & -42.5020 & 0.4828 & 0.0922 & 33.3084 & -30.8538 \\
2023 & 1.2883 & 0.1069 & 11.4713 & -9.0995 & 0.4282 & 0.1792 & 16.3157 & -13.0680 \\
\bottomrule
\end{tabular}
     
\caption{Average parameters over the different maturities for Germany in January 2021, 2022 and 2023, calibrated over full trading sessions with normalised time from ~\eqref{eq:min_param_time}. For the $\alpha$ parameters, we multiply them by the duration of the trading session (which is equivalent to rescaling the time by $T_m$) to give them the same order of magnitude before averaging them. For example, to apply these parameters specifically to the 8-hour (8H) product, the parameter $\alpha$ given in this table should be divided by a factor of $8+9 = 17$. Here the extra $9$ hours represent the time between the opening of the session and midnight.
}
\label{tab:params_liq_germany}
\end{table}

\begin{table}[H]
\centering
        \begin{tabular}{rrrrrrrrr}
\toprule
Year & $\alpha_{A_{m,+}} T_m$ & $\beta_{A_{m,+}}$ & $\alpha_{B_{m,+}}T_m$ & $\beta_{B_{m,+}}$ & $\alpha_{A_{m,-}}T_m$ & $\beta_{A_{m,-}}$ & $\alpha_{B_{m,-}}T_m$ & $\beta_{B_{m,-}}$ \\
\midrule
2021 & 0.1854 & 0.2517 & 3.2813 & -0.6146 & 0.1468 & 0.2746 & 3.5512 & -0.7842 \\
2022 & 1.8613 & 0.7615 & 5.7148 & -1.3478 & 1.5962 & 0.6560 & 3.3275 & 0.4182 \\
2023 & 0.3440 & 0.6388 & 7.8779 & -3.0430 & 0.6829 & 0.4090 & 7.6798 & -2.8904 \\
\bottomrule
\end{tabular}
     
\caption{Same as Table \ref{tab:params_liq_germany}, for France.
}
\label{tab:params_liq_france}
\end{table}
%Note that using \eqref{eq:min_params} over the whole trading session can introduce bias, as the majority of data is concentrated toward the end of the session when liquidity is highest. This can result in a model that poorly represents the early stages of the trading session. To address this issue, one approach is to apply weights to the criterion~\eqref{eq:min_params}. 

\subsection{Mid-price modelling}
\label{sec:modelling_price}

In this section we use the model of Deschatre and Warin~\cite{deschatre2024} for the mid-price modelling part and remind some results from~\cite{deschatre2024}. The model is constructed from three-dimensional Poisson measures, but for simplicity we give the construction from compound Poisson processes of~\cite[Corollary 3.1]{deschatre2024}. On a probability space $\left(\Omega, \mathcal{F}, \mathbb{P}\right)$, with $\mu$, $\mu_c$ and $\kappa > 0$, let $P^{+}_1,\,P^{-}_1,\ldots, P^{+}_M,\,P^{-}_M,\,P^{c,+}_1,\,P^{c,-}_1,\ldots,P^{c,+}_M,\,P^{c,-}_M$ be $4M$ independent compound Poisson processes with intensities of 
\begin{itemize}
    \item[-] $\mu e^{-\kappa(T_m-s)}$ for $P^{+}_m$ and $P^{-}_m$, $m \geq 1$,
    \item[-] $\mu_c e^{-\kappa(T_M-s)}$ for $P^{c,+}_M$ and $P^{c,-}_M$,
    \item[-] $\mu_c e^{-\kappa(T_{m}-s)} - \mu_c e^{-\kappa(T_{m+1}-s)}$ for $P^{c,+}_m$ and $P^{c,-}_m$, $m=1,\ldots,M-1$,
    \end{itemize}
and jump law $\nu(dy)$ on a measurable space $(K,\mathcal{K})$, $K \subset \mathbb{R}_+$, with $\nu(\{0\})=0$. The filtration considered in the following $(\mathcal{F}_t)_t$ is the natural filtration of all these compound Poisson processes.

\medskip
If $f_{m,0}$ is the initial price for the maturity $T_m$, then the price for the maturity $T_m$ is given by ($x \wedge y$ is the notation for the minimum between $x$ and $y$)
\begin{equation}
\label{eq:model}    
f_{m,t} = f_{m,0} + f_{m,t\wedge T_m}^{+} - f_{m,t\wedge T_m}^{-},
\end{equation}
with
\begin{equation} \label{eq:model_pm}
f^h_{m,t} = P^{h}_{m,t}+ \sum_{j=m}^M P^{c,h}_{j,t}
\end{equation}
for $h=+,-$, $m=1,\ldots,M$ and $0 \leq t \leq T$. $f^+_{m,t}$ is the sum of the positive jumps of the process up to time $t$ for maturity $M$, $f^-_{m,t}$ is the sum of the negative ones. Each of these processes is the sum of a compound Poisson process $P^{h}_{m,t \wedge T_m}$ specific to the maturity $m$ and of the jump process $\sum_{j=m}^M P^{c,h}_{j,t \wedge T_m}$ correlating the different maturities. The last term represents a common shock occurring in the market and affecting all maturities (e.g. power plant failure, increase in temperature forecast): a jump occurring in $P^{c,h}_k$   affects the prices with maturity $T_1$, \ldots $T_k$ simultaneously and in the same direction. Common Shock modelling is a standard way to create correlation between Poisson processes~\cite{powojowski2002, lindskog2003}. The price process $(f_m)_{m\in \{1,\ldots,M\}}$ is a quadratic integrable $(\mathcal{F}_t)_t$ martingale from \cite[Proposition 1]{deschatre2024}.

\medskip
If we look at $f^h_{m,t}$, it is easy to see that it is a compound Poisson process with jump distribution $\nu$ and intensity
\[
\mu e^{-\kappa(T_m-s)} + \sum_{j=m}^{M-1} (\mu_c e^{-\kappa(T_{j}-s)} - \mu_c e^{-\kappa(T_{j+1}-s)}) +\mu_c e^{-\kappa(T_{M}-s)}  = (\mu + \mu_c) e^{-\kappa(T_m-s)}.
\]
The intensity of the price changes for the maturity $m$ is the intensity of the jumps of $f^+_{m,t} + f^-_{m,t}$, i.e. 
\begin{equation} \label{eq:intensity}
2 (\mu + \mu_c) e^{-\kappa(T_m-s)}
\end{equation}
and increases with time to maturity, which is consistent with the empirical findings on mid-prices of \cite{deschatre2023b, bergault2024}. A proxy for the integrated volatility up to time $t$ for the price process associated with the maturity $m$, computed from the expectation of the quadratic variation, is $\int_0^{t \wedge T_m} \sigma_s^2 ds$ with
\begin{equation} \label{eq:vol_th}
    \sigma^2_{m,t} = 2\int_K y^2 \nu(dy) \left(\mu + \mu_c\right)e^{-\kappa \left(T_m - s\right)},
\end{equation}
see \cite[Proposition 3]{deschatre2024} and the discussion below. The volatility parameter increases as time approaches maturity, as does the intensity of price changes: this is the so-called Samuelson effect and is consistent with the mid-price data, see~\cite{deschatre2023b}.

\begin{figure}[H]
    \centering
    \begin{tabular}{cc}
       \includegraphics[width=0.45\textwidth]{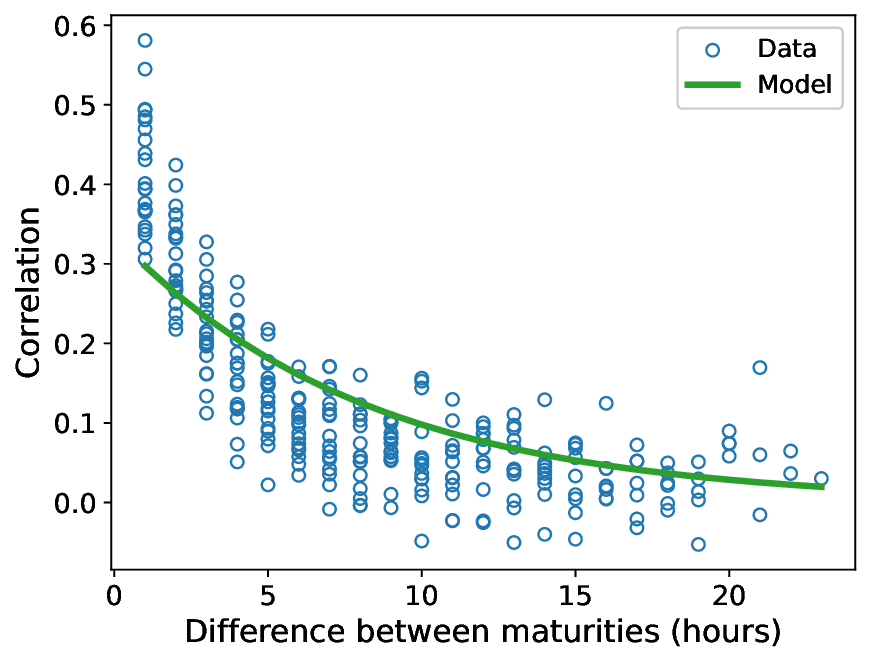}  &  \includegraphics[width=0.45\textwidth]{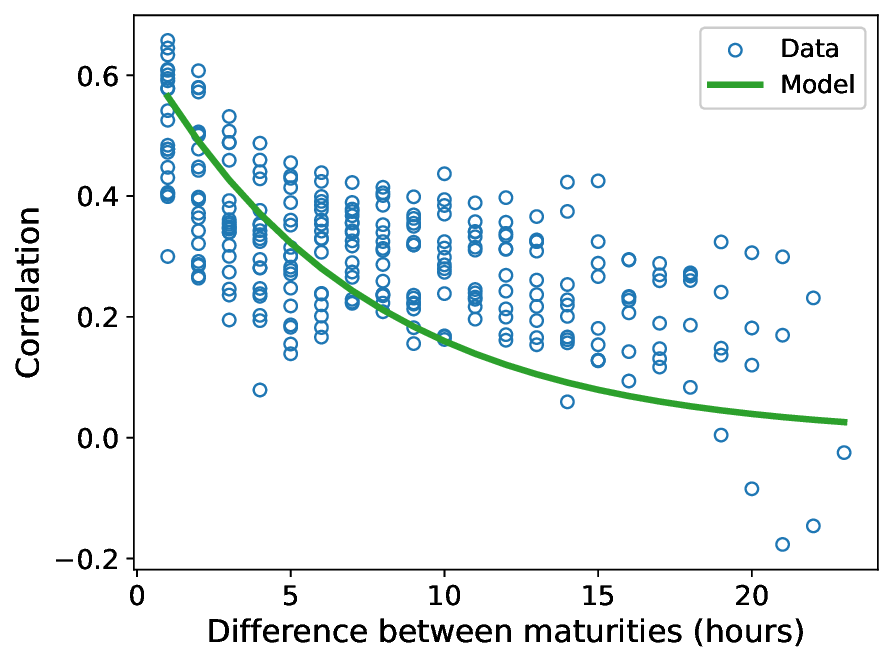}\\
       (A) Germany & (B) France
    \end{tabular}
    \caption{Correlations as a function of maturity distance estimated using the empirical counterpart of quadratic covariation with a sampling time step of 30 minutes (blue dots) for German (left) and French (right) mid-prices in January 2021. The green line corresponds to the model correlation computed in Equation~\eqref{eq:correl_th}.}
    \label{fig:reg_correl}
\end{figure}

\medskip
In the same way, we can compute a proxy for the correlation from the quadratic covariation. The correlation between $f_k$ and $f_l$ for $k \neq l$ does not depend on $t$ and is given for $t \leq \min(T_k,T_l)$ by 
\begin{equation} \label{eq:correl_th}
\rho_{lk} = \frac{\mu_c}{\mu + \mu_c}e^{-\frac{\kappa}{2} \left|T_l - T_m\right|}.
\end{equation} 
The correlation decreases with the distance between the two maturities: this is called the Samuelson \textit{correlation} effect. It has just been identified empirically for transaction prices in \cite{deschatre2024, hirsch2024b}. It also holds for the mid-prices in France and Germany, see Figure~\ref{fig:reg_correl}.

\medskip
This model requires only three parameters in addition to the law of jumps:
\begin{itemize}
    \item[(i)] $\kappa > 0$ is the rate of increase of the intensity of mid-price changes given by Equation~\eqref{eq:intensity} and of the volatility given by Equation~\eqref{eq:vol_th}. $\kappa/2$ is also the rate at which the correlation between two maturities decreases, given by Equation~\eqref{eq:correl_th}, as the distance between these maturities increases;
    \item[(ii)] $\mu > 0$ is a measure of the intensity of mid-price movements that occur independently at each maturity;
    \item[(iii)] $\mu_c > 0$ is a measure of the intensity of shocks that affect several maturities simultaneously.
\end{itemize}

\medskip
Finally, note that a diffusion model proxy for this model (obtained with for large $\mu$ and $\mu_c$) is
\[
\left(\int_0^{t \wedge T_m} \sigma_{m,s} dW_{m,s}\right)_{m=1,\ldots,M},\,t \in \left[0, T\right]
\]
where $W = \left(W_1, \ldots,W_M\right)^{\top}$ is a multivariate Brownian motion with correlation matrix $\rho_{lk}$ given by Equation~\eqref{eq:correl_th} and $\sigma^2_{m,t}$ is given by Equation~\eqref{eq:vol_th}, see \cite[Proposition 6]{deschatre2024}.

\medskip
The estimation procedure is a moment-based method, with $\kappa$ estimated from the intensity of the mid-price moves, and $\mu$ and $\mu_c$ estimated from empirical quadratic variations and covariations, with a time step $\Delta$ large enough to remove microstructure noise, which we choose to be 30 minutes. Note also that we consider the returns that are greater in absolute value than 5 times the standard deviation of all non-zero returns over the estimation period as outliers and remove them as in~\cite{deschatre2024}. We give the estimated parameters for the three estimation periods January 2021, January 2022, and January 2023 for Germany and France respectively in Table~\ref{tab:params_germany} and Table~\ref{tab:params_french}, as well as empirical estimators of the two first moments of the jump sizes. The empirical distribution for the law of jump sizes is used in simulations. The intensities $\mu$ and $\mu_c$ are much higher for Germany: the mid-prices move with a higher frequency. The size of the jumps is much larger for the French market, indicating a sparser order book. This is consistent with the German market being more liquid than the French market. The high values of the jump sizes make the proxy parameter for the squared integrated variance $\sigma=\underset{T_m \to \infty}{\lim}\sqrt{\int_0^{T_m} \sigma_{m,s}^2ds}$ higher for France than for Germany. We also notice an increase in this proxy parameter in 2022, which corresponds to a year of high volatility due to the war in Ukraine (which affected gas prices) and some nuclear plant shutdowns at the end of 2021 in France. The correlation parameter $\rho_1$ between two consecutive products is of the same order of magnitude for the two countries and the different years and seems to be underestimated for Germany by our model if we look at Figure~\ref{fig:reg_correl}. 

\begin{table}[H]
\centering
        \begin{tabular}{rrrrrrrr}
\toprule
Year & $\kappa$ & $\mu$ & $\mu_c$ & $\int_K y \nu(dy)$ & $\int_K y^2 \nu(dy)$ & $\sigma$ & $\rho_1$ \\
\midrule
2021 & 0.25 & 109.45 & 55.45 & 0.09 & 0.04 & 7.60 & 0.26 \\
2022 & 0.28 & 195.12 & 214.02 & 0.22 & 0.58 & 41.51 & 0.40 \\
2023 & 0.23 & 181.09 & 172.83 & 0.13 & 0.21 & 25.20 & 0.39 \\
\bottomrule
\end{tabular}
     
\caption{Estimated parameters for the model~\eqref{eq:model}-\eqref{eq:model_pm} on German mid-prices for the years 2021, 2022 and 2023 and month January. The unit for $\kappa$, $\mu$ and $\mu_c$ is the inverse of the hour. $\sigma=\underset{T_m \to \infty}{\lim}\sqrt{\int_0^{T_m} \sigma_{m,s}^2ds}$ is a proxy for the squared integrated volatility where $\sigma^2_{m,s}$ is given in Equation~\eqref{eq:vol_th}. $\rho_1$ is the correlation between two consecutive products ($\rho_{ij}$ of Equation~\eqref{eq:correl_th} with $|i-j|=1$).
}
\label{tab:params_germany}
\end{table}

\begin{table}[H]
\centering
        \begin{tabular}{rrrrrrrr}
\toprule
Year & $\kappa$ & $\mu$ & $\mu_c$ & $\int_K y \nu(dy)$ & $\int_K y^2 \nu(dy)$ & $\sigma$ & $\rho_1$ \\
\midrule
2021 & 0.28 & 11.50 & 21.33 & 0.32 & 1.28 & 17.32 & 0.49 \\
2022 & 0.36 & 31.90 & 34.23 & 0.83 & 6.75 & 49.55 & 0.36 \\
2023 & 0.19 & 20.78 & 42.19 & 0.35 & 3.11 & 45.77 & 0.56 \\
\bottomrule
\end{tabular}
     
\caption{Same as Table~\ref{tab:params_germany}, for France.
}
\label{tab:params_french}
\end{table}

\section{Numerical results}
\label{sec:optim}
\subsection{Optimizing position on one index}

In \cite{deschatre2024}  it has been shown that the proposed model is a good candidate for valuing a battery in the intraday market and gives a good strategy that outperforms in backtesting deterministic classical strategies based on the hourly spot values taken as a perfect forecast of the intra-day prices of the same delivery periods. Without taking into account the liquidity of the market, the previous result is only valid when there are a small number of batteries in the market.

\medskip
In this section we want to show that taking into account the liquidity of the market is very relevant for a good valuation of batteries, especially when there are a large number of batteries in the market. Moreover, since the market is imperfect, it is also relevant to decide which index to use to take decisions: if a decision concerns the management of the battery at a given hour $H$, the decision must be taken at the time $H-\Delta$, where $\Delta$ should be between $1$ and a few hours, since the market is illiquid. We consider the case of a 2h battery and the case of a 3h battery. A $n$h battery has the following characteristics:
\begin{itemize}
    \item[-] the capacity of the battery is $n$~MWh;
    \item[-] the injection and withdrawal capacities are $1$MWh per hour.
\end{itemize}
The battery efficiency is assumed to be $\rho=0.92$, so:
\begin{itemize}
    \item[-] putting $E$ MWh into storage requires us to take $\frac{E}{\rho}$ MWh from the grid, so we buy the equivalent amount on the market;
    \item[-] withdrawing $E$ MWh from storage only adds $\rho E$ MWh to the grid, and is therefore equivalent to selling that amount on the market.
\end{itemize}
As in \cite{deschatre2024}, we consider the intraday prices $f_{m,t}$ for the delivery period $\left[T_m, T_m+\theta\right]$ with $\theta = 1$ hour, with $T_1=0,\ldots,T_{24}=23$ (every hour of the day). Every hour a decision is made with a delay of $\Delta =1$ or $\Delta=2$ hours: the decision is therefore based on the price $f_{m, T_{m-\Delta}}$ for each $m \in \{1,\ldots,24\}$. The battery is managed assuming zero inventory at $T_1=0$ hour each day. The objective function is to maximize the expected profit at $T_0=15$h on the day before management, which is the opening of the intraday market for the maturities under consideration. Taking into account the liquidity model, the purchase of a volume $V$ (positively counted if purchase) at date $T_m - \Delta$ for delivery at date $T_m$, costs a price $P$ per unit of volume: 
\begin{align*}
     P( m, \Delta,  V) = & f_{m, T_{m-\Delta}} +  [A_{m,-}(T_m-\Delta) V - B_{m,-}(T_m-\Delta)]{\bf 1}_{\{V<0\}} +  \\
     & [A_{m,+}(T_m-\Delta)V + B_{m,+}(T_m-\Delta)]{\bf 1}_{\{V > 0\}}.
\end{align*}
The non-anticipative control taken at the time $T_i -\Delta$ belonging to $\mathcal{F}_{T_i-\Delta } = \{ f_{m, s} | s \le T_i -\Delta,\, m=1,\ldots,24\}$ is noted $C_i$ (a positive $C_i$ corresponds to an injection) and we note $\tilde C= (C_1, \dots, C_{24})$. For a number of batteries $\hat N$ present, and since all batteries are identical, the value function of a single battery is obtained by optimizing $J$:
\begin{align}
\label{eq:optSto}
    J(\tilde C) =  -\E[  \sum_{i=1}^{24}  C_i ( \frac{1}{\rho} 1_{C_i \ge 0} + \rho 1_{C_i \le 0} ) P( m, \Delta, \hat N C_i) | \mathcal{F}_{T_0}]
\end{align}
with the constraints for $i=1, \dots, 24$:
\begin{align*}
    0 \le & \sum_{j=1}^i C_j  \le \bar C, \\
    - \underline{C}&  \le C_i   \le \underline{C} .
\end{align*}
In our case, $\bar C= n$ and $\underline C= 1$.

\medskip
In \cite{deschatre2024}, it was shown that it is possible to accurately estimate conditional expectations at hour $T_m$, keeping only information on $4$ products of the shortest maturity $(f_{m+\Delta +i,T_{m}})_{i=0,3}$, using regressions with 
 local adaptive linear bases from \cite{bouchard2012monte} in the StOpt library \cite{gevret2018stochastic}: $500000$ Monte Carlo price trajectories and $4$ meshes in each dimension for $p \le 4$, and one mesh per dimension beyond 4, giving us a total of $4^{p \wedge 4}$ meshes, are used to estimate conditional expectations optimizing \eqref{eq:optSto}.
 Moreover, due to the impact model, the problem \eqref{eq:optSto} is no longer linear: the control is no longer bang-bang, and using dynamic programming it is necessary to discretize the stocks and commands thinly (see \cite{warin2012}). In our test, all stocks and commands are discretized with a step of $0.1$ MWh.

\medskip
To use our model on a day $D$ on a given market with a given $\Delta$, we estimate the price model and the liquidity parameters using the last 28 days of data and parameters are updated every Monday of each week. For each day $D$ of the year, different stochastic optimizations are obtained:
\begin{itemize}
     \item[-] a first (Stochastic no depth model) solved \eqref{eq:optSto} with the parameters set for the market under study on the current day, but without price impact, thus replacing $P( m, \Delta, \hat N C_i)$ by $f_{i,T_{i} - \Delta} $, giving an optimal intraday storage management strategy for the model used, assuming no price impact.
     \item[-] a second (Stochastic depth model) solves \eqref{eq:optSto} taking into account the impact model for a given number of batteries $\hat N$ settled in the market, giving an optimal strategy for the model used assuming price impact and $\hat N$ given.
\end{itemize}
These strategies can be compared to a ``spot control'' optimisation strategy with or without price impact : in these strategies, the control is calculated from the spot prices $\{ f_{i,T_0} \}_{i=1,\ldots, 24}$ at $T_0$.
The optimal control $\tilde D= (D_1, \dots, D_{24})$ is obtained by maximising the following problem:
\begin{align}
     \hat J( \tilde D) = - \sum_{i=1}^{24}  D_i ( \frac{1}{\rho} 1_{D_i \ge 0} + \rho 1_{D_i \le 0} ) \hat P( m, \Delta, \hat N C_i), \label{eq:optDet}
 \end{align}
 where
 \begin{align*}
    \hat P( m , \Delta, V) =  & f_{m, T_{0}} -  [A_{m,-}(T_m-\Delta) V - B_{m,-}(T_m-\Delta)]{\bf 1}_{\{V<0\}} + [A_{m,+}(T_m-\Delta)V +  \\
    & B_{m,+}(T_m-\Delta)]{\bf 1}_{\{V > 0\}}
 \end{align*}
 considering the price impact, or $\hat P( m,   V) =  f_{m,T_0}$ if no price impact is considered.
 The Equation \eqref{eq:optDet} is solved 
under the constraints for $i=1, \dots, 24$
 \begin{align*}
     0 & \le \sum_{j=1}^i D_j  \le \bar C, \\
     - \underline{C} & \le D_i   \le \underline{C} .
 \end{align*}
So for the same day $\hat D$, two deterministic optimisations are achieved:
\begin{itemize}
     \item[-] A first  (Deterministic no depth model) solves \eqref{eq:optDet} without price impact, giving a first deterministic strategy independent of the number of batteries $\hat N$.
     \item[-] A second (Deterministic depth model) solves \eqref{eq:optDet} with price impact giving a deterministic strategy depending on $\hat N$.
\end{itemize}

\medskip
The four strategies calculated (2 stochastic and 2 deterministic) for the given day $\hat D$ can be tested in a back test using the available order book. Then, at each hour $h$ of the day $\hat D$, the calculated strategies give us the volume of the product of maturity $h+ \Delta$ to buy or sell and using the order book we get the exact cost or gain associated to our control.
Therefore for a given $\hat D$ we get a profit associated to each strategy for a given $\hat N$ and we can get for each year the real profit that we could have obtained.

\medskip
The results in Table~\ref{tab:fr2021_combined} and Table~\ref{tab:al2021_combined} for the year 2021 indicate that initiating positions one hour before delivery is suboptimal. This is primarily because the XBID market closes one hour prior to delivery, prompting traders to finalize their trades just before the closure. Consequently, a gradual decrease in liquidity is observed between $1.5$ hours and $1$ hour before maturity. This trend highlights the advantage of setting $\Delta=2$, which leads to improved outcomes. Moreover, on both markets, the results with either a $2h$ or a $3h$ battery behave in the same way in 2021. 
In the years 2022 and 2023 we only provide in the Appendix~\ref{sec:appendix} results for a $2h$ battery with $\Delta=2$ hours in Table~\ref{tab:fr2022_2023_combined} and Table~\ref{tab:al2022_2023_combined} respectively for the France and the Germany. The liquidity in France is rather low and even with a single battery, there is a small gain when taking into account the price impact as shown on Table~\ref{tab:fr2021_combined}. Going up to $20$ batteries, as the price impact model has been fitted with a window depth of $20$ MW, not taking into account the market impact can lead to heavy losses: we expect the profit with $N$ batteries to be higher than the profit with $M$ batteries if $M < N$ : this is clearly not the case if we don't take into account the price impact in 2021. On the more liquid German market, the price impact model is fitted with a depth of $100$ MW in Table~\ref{tab:al2021_combined}.
As expected for a single battery, the price impact model does not improve backtesting profits.
Up to 20 batteries, the price impact model does not bring much more profit. In 2021, even with 50 batteries, the gains are small.

\begin{table}[H]
    \centering
    \begin{tabular}{rrr|rr|rr|rr}
    \toprule
    $n$ & $\Delta$ &  $\hat{N}$ & \multicolumn{2}{c}{1} & \multicolumn{2}{c}{10} & \multicolumn{2}{c}{20} \\
    & & Model & Depth & No depth & Depth & No depth & Depth & No depth \\
    \midrule
    \multirow{4}{*}{2} & \multirow{2}{*}{1} & Det. 
    & 21264 & 18657 & 11737 & 5718 & -590 & -17280 \\
    &  & Sto. 
    & 25965 & 23994 & 18446 & 9353 & 10954 & -14248 \\
    \cline{2-9}
    & \multirow{2}{*}{2} & Det. 
    & 32152 & 31919 & 26772 & 24954 & 17579 & 13185 \\
    &  & Sto. 
    & 33914 & 33699 & 29757 & 26263 & 22504 & 11872 \\
    \midrule
    \multirow{2}{*}{3} & \multirow{2}{*}{2} & Det. 
    & 42612 & 41961 & 34287 & 31472 & 21318 & 13902 \\
    &  & Sto. 
    & 45397 & 44982 & 38305 & 35509 & 27901 & 17076 \\
    \bottomrule
    \end{tabular}
    \caption{Backtest gain (euros per battery) for the optimisation of $\hat{N}$ $n$h batteries for the year 2021 on the French market for the different models. The market position is taken $\Delta$ hours before maturity. Det. is for Determinist and Sto. for Stochastic.}
    \label{tab:fr2021_combined}
\end{table}

\begin{table}[H]
    \centering
    \begin{tabular}{rrr|rr|rr|rr|rr}
    \toprule
    $n$ & $\Delta$ & $\hat{N}$ & \multicolumn{2}{c}{1} & \multicolumn{2}{c}{20} & \multicolumn{2}{c}{50} & \multicolumn{2}{c}{100} \\
    & &Model  & Depth & No depth & Depth & No depth & Depth & No depth & Depth & No depth \\
    \midrule
    \multirow{4}{*}{2} & \multirow{2}{*}{1} & Det.
    & 32178 & 32060 & 27672 & 26287 & 11649 & 5353 & -24272 & -84059 \\
    &  & Sto.
    & 44727 & 44720 & 40535 & 39439 & 32318 & 23416 & 18481 & -38972 \\
    \cline{3-11}
    & \multirow{2}{*}{2} & Det.
    & 35584 & 35536 & 34287 & 33626 & 29450 & 27492 & 12011 & 4495 \\
    &  & Sto.
    & 45500 & 45585 & 43711 & 43658 & 39760 & 38000 & 31543 & 17689 \\
    \midrule
    \multirow{2}{*}{3} & \multirow{2}{*}{2} & Det.
    & 48891 & 48854 & 47043 & 46113 & 38874 & 36695 & 14079 & 3713 \\
    &  & Sto.
    & 61752 & 61791 & 59297 & 59152 & 53313 & 51204 & 37295 & 21108 \\
    \bottomrule
    \end{tabular}
    \caption{Backtest gain (euros per battery) for the optimisation of $\hat{N}$ $n$h batteries for the year 2021 on the German market for the different models. The market position is taken $\Delta$ hours before maturity. Det. is for Determinist and Sto. for Stochastic.}
    \label{tab:al2021_combined}
\end{table}

\begin{figure}[H]
     \centering
     \includegraphics[width=0.5\linewidth]{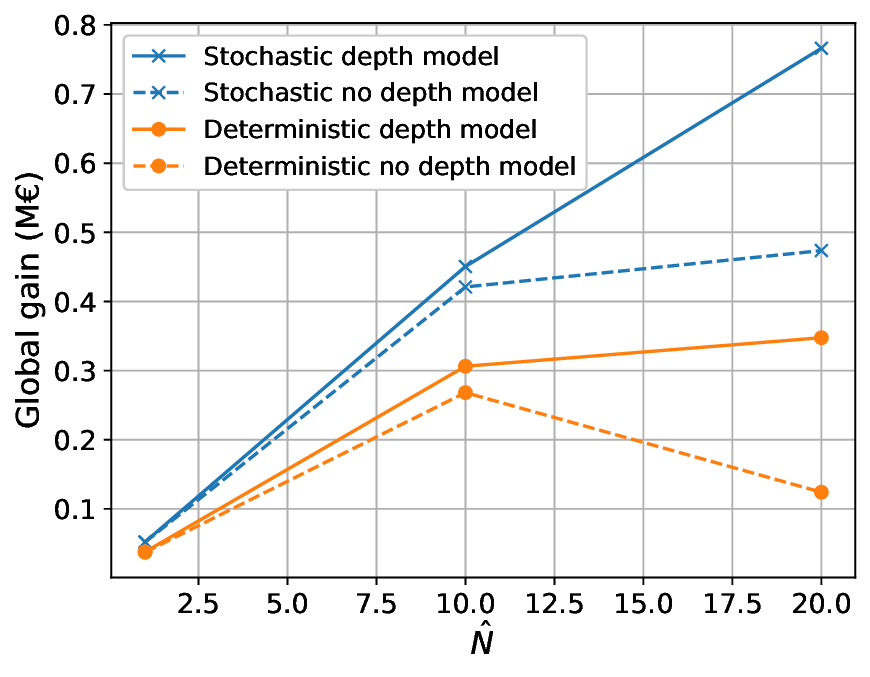}
     \caption{Global gain in millions of euros for a park of $\hat N$  $2h$ batteries in France in 2023.}
     \label{fig:globGainFR2023}
 \end{figure}

\begin{figure}[H]
    \centering
    \begin{tabular}{cc}
      \includegraphics[width=0.45\linewidth]{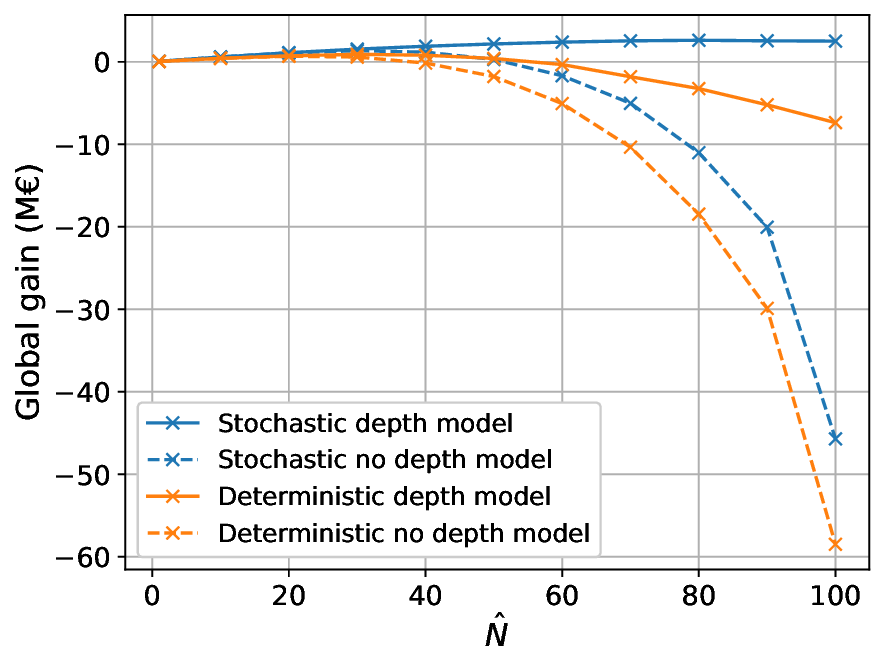} & 
      \includegraphics[width=0.45\linewidth]{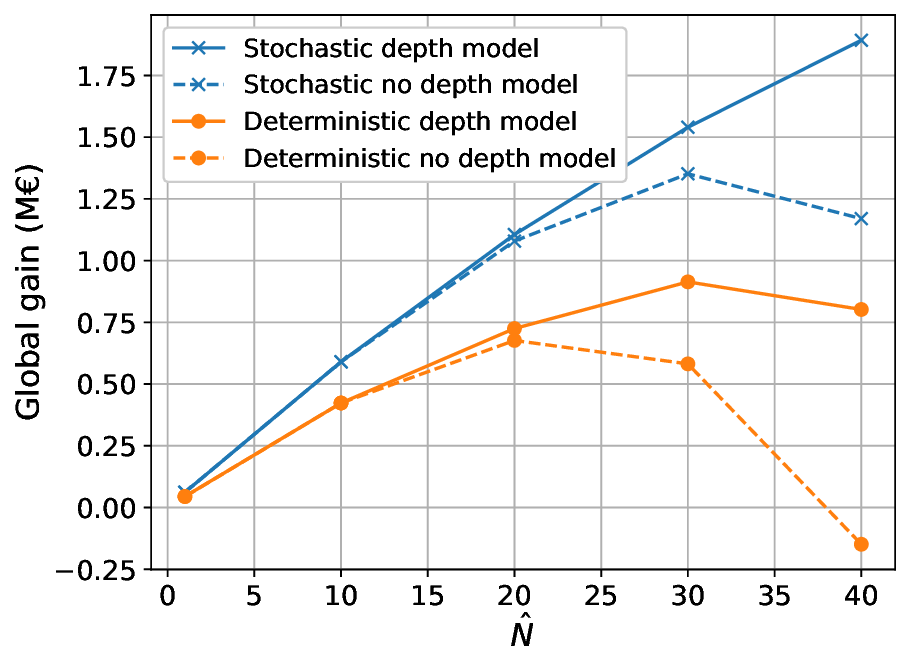} \\
      (A) Results for $\hat N \in [1,100]$ & (B) Zoom for $\hat N \in [1,40]$
    \end{tabular}

    \caption{Global gain in millions of euros for a park of $\hat N$  $2h$ batteries in Germany in 2023.}
    \label{fig:globGainAL2023}
\end{figure}

In Figure~\ref{fig:globGainFR2023} and Figure~\ref{fig:globGainAL2023}, we show the global gain of a park of $\hat N$ batteries achieved in the backtest in 2023. When the number of batteries is below $20$ in France and $40$ in Germany, the gain is visually almost linear with the number of batteries, taking into account the depth. For Germany, the losses without taking the depth into account are huge for a number of batteries above $60$.
 
Taking liquidity into account is of utmost importance, especially when the number of batteries is large: it enables much greater gains to be made and avoids losses. The use of a stochastic model, as already shown in~\cite{deschatre2024}, makes it possible to improve the results without taking liquidity into account; here we also show that with taking liquidity into account, the results are much better.

\subsection{Optimizing on two indexes}
In the previous section, we have studied the management of a battery taking a decision one ($\Delta=1$) or two hours ($\Delta=2$) before maturity. One may wonder if it is possible to take into account the dynamics of the index for a given maturity : thus, for a given date, we could take a position for products with delivery in $2$ and $3$ hours. Since the pricing model used without price impact is a martingale, there is theoretically no interest in taking a position on two indexes at a given hour if no price impact is taken into account.
The possibility to trade two indexes is only a way with our model to have more liquidity and splitting a volume traded on both hours it could bring us an advantage. We can try to optimize the previous problem by making 2 decisions at each hour and backtesting our strategy. This problem is much harder to solve because it is an optimization problem with 2 stocks: the first stock corresponds to the position taken on the product with a maturity of 3 hours, while the second corresponds to the additional volume traded 2 hours before maturity. In Table~\ref{tab:index2FR_AL_combined}, we optimize a storage on a day at the beginning of 2021 with the stochastic price model, taking into account the market impact on the French and German markets, and report the theoretical  battery profit as a function of $\hat N$. As expected, the profit of trading the two indexes is higher than that of trading the 1 index, as it is a special case of the two indexes problem taking 0 volume trading product with a maturity of 3 hours.

\begin{comment}
\begin{table}[H]
    \centering
    \begin{tabular}{ccc}
    \toprule
     Number of batteries    &  Gain one index &  Gain two index\\
     \midrule
       1   & 129 &153 \\  
      10   & 108 &129 \\  
      20   & 91 &  111\\  
        \bottomrule

    \end{tabular}
    \caption{France, $2h$ battery ,  theoretical gain per battery with the model trading  products $T_{m}+2$, $T_{m}+3$}
    \label{tab:index2FR}
\end{table}
\begin{table}[H]
    \centering
    \begin{tabular}{ccc}
    \toprule
     Number of batteries    &  Gain one index &  Gain two index\\
\midrule
       1   &  113 & 120 \\  
      10   & 106 & 112 \\  
      20   & 100 & 107  \\  
      50   &  84  & 95 \\  
\bottomrule
    \end{tabular}
    \caption{Germany, $2h$ battery, theoretical  gain per battery with the model trading  products $T_{m}+2$, $T_{m}+3$}
    \label{tab:index2AL}
\end{table}
\end{comment}

\begin{table}[H]
    \centering
    \begin{tabular}{cccc}
    \toprule
    Country & $\hat{N}$ & Gain one index & Gain two indexes \\
    \midrule
    France  & 1  & 129 & 153 \\
            & 10 & 108 & 129 \\
            & 20 & 91  & 111 \\
    \midrule
    Germany & 1  & 113 & 120 \\
            & 10 & 106 & 112 \\
            & 20 & 100 & 107 \\
            & 50 & 84  & 95  \\
    \bottomrule
    \end{tabular}
    \caption{Theoretical gain (euros per battery) for $\hat{N}$ $2h$ battery with the model trading products $T_{m}+2$, $T_{m}+3$ (gain two indexes) versus trading only product $T_m+2$ (gain one index) for France and Germany with parameters estimated on the 28th of december 2020.}
    \label{tab:index2FR_AL_combined}
    \end{table}

Backtesting the two indexes  strategy for a full year is computationally too expensive. Therefore, we backtest it only for the first 50 days of 2021 with $\hat N=1$ and give results in Table~\ref{tab:backtest2index}. We can clearly see that the backtesting of two indices results in a significant loss compared to the optimisation of a single index.
The pricing model used is not realistic enough to take into account the correlations between the index prices.

\begin{table}[H]
    \centering
    \begin{tabular}{ccc}
    \toprule
     Market  &  Gain one index &  Gain two indexes\\
     \midrule
       France   &  2170 & 487\\  
       Germany   & 2269  & 1620 \\  
\bottomrule
    \end{tabular}
    \caption{Backtest gain (euros per battery) for one $2h$ battery with the model trading products $T_{m}+2$, $T_{m}+3$ (gain two indexes) versus trading only product $T_m+2$ (gain one index) for France and Germany on the first 50 days of 2021.}
    \label{tab:backtest2index}
\end{table}

\vip 
\section*{Acknowledgements}

The authors acknowledge support from the FiME Lab (Institut Europlace de Finance). The authors are grateful to F\'elix Trieu for his valuable help with the data and to Pierre Gruet for his constructive comments.

\section*{Disclosure statement}

The authors report there are no competing interests to declare.
%********************** Bibliography
\bibliographystyle{plain}
\bibliography{biblio}
%\printbibliography
\appendix

\section{Results for years 2022 and 2023}
\label{sec:appendix}
In this appendix we give for years 2022 and 2023 the different values obtained in back-test for a single battery of a park of $\hat{N}$ batteries in Table~\ref{tab:fr2022_2023_combined} for the French market and Table~\ref{tab:al2022_2023_combined} for the German market.

\begin{comment}
\begin{table}[H]
    \centering
    \begin{tabular}{c|c|c|c|c|c|c}
       $\hat{ N}$   & \multicolumn{2}{c}{1} &  \multicolumn{2}{c}{10} & \multicolumn{2}{c}{20} \\
       Model        & Depth & No depth& Depth & No depth& Depth & No depth \\
       Deterministic&  68246   &  67263  &  59922 & 56031 & 42707 &  34959  \\
       Stochastic & 76553 & 76570  &  68749 & 66480 &  60020 &   48130
    \end{tabular}
    \caption{France,$2h$ battery, $\Delta =2$ : backtest gain on a single battery in 2022}
    \label{tab:fr2022Idec2}
\end{table}
\begin{table}[H]
    \centering
    \begin{tabular}{c|c|c|c|c|c|c}
       $\hat{ N}$   & \multicolumn{2}{c}{1} &  \multicolumn{2}{c}{10} & \multicolumn{2}{c}{20} \\
       Model        & Depth & No depth& Depth & No depth& Depth & No depth \\
      Deterministic&   37785  &  37245  & 30626  & 26815 & 17378 &  6204  \\
       Stochastic & 52122 & 51890  & 45052  & 42100 & 38311  & 23679  
     \end{tabular}
    \caption{France,$2h$ battery, $\Delta =2$ : backtest gain on a single battery in 2023}
    \label{tab:fr2023Idec2}
\end{table}
\end{comment}
\begin{table}[H]
    \centering
    \begin{tabular}{rr|rr|rr|rr}
    \toprule
    Year &  $\hat{N}$ & \multicolumn{2}{c}{1} & \multicolumn{2}{c}{10} & \multicolumn{2}{c}{20} \\
    &Model & Depth & No depth & Depth & No depth & Depth & No depth \\
    \midrule
    \multirow{2}{*}{2022} & Det. 
    & 68246 & 67263 & 59922 & 56031 & 42707 & 34959 \\
    & Sto. 
    & 76553 & 76570 & 68749 & 66480 & 60020 & 48130 \\
    \midrule
    \multirow{2}{*}{2023} & Det. 
    & 37785 & 37245 & 30626 & 26815 & 17378 & 6204 \\
    & Sto. 
    & 52122 & 51890 & 45052 & 42100 & 38311 & 23679 \\
    \bottomrule
    \end{tabular}
    \caption{Backtest gain (euros per battery) for the optimisation of $\hat{N}$ $2$h batteries for the years 2022 and 2023 on the French market for the different models. The market position is taken $2$ hour before maturity. Det. is for Determinist and Sto. for Stochastic.}
    \label{tab:fr2022_2023_combined}
\end{table}

\begin{comment}
\begin{table}[H]
    \centering
    \begin{tabular}{c|c|c|c|c|c|c|c|c}
       $\hat{ N}$   & \multicolumn{2}{c}{1} &  \multicolumn{2}{c}{20} & \multicolumn{2}{c}{50}& \multicolumn{2}{c}{100} \\
       Model        & Depth & No depth& Depth & No depth& Depth & No depth& Depth & No depth \\
       Deterministic&  86591   &  86605& 140591&   80845 & 79625  & 56381& -18218 &  -101280  \\
       Stochastic & 106882 & 106813  & 101778& 101284&   90521& 79625 & 68123  &   -56460
    \end{tabular}
    \caption{Allemagne, $2h$ battery, $\Delta =2$ : backtest gain on a single battery in 2022}
    \label{tab:al2022Idec2}
\end{table}

\begin{table}[H]
    \centering
    \begin{tabular}{c|c|c|c|c|c|c|c|c}
       $\hat{ N}$   & \multicolumn{2}{c}{1} &  \multicolumn{2}{c}{20} & \multicolumn{2}{c}{50}& \multicolumn{2}{c}{100} \\
       Model        & Depth & No depth& Depth & No depth& Depth & No depth& Depth & No depth \\
       Deterministic&   44812  & 44416   & 36246 & 33833 &  8164 & -35606&  -73747 &  -584819  \\
       Stochastic & 61823 & 61824  & 55285  &  53964 & 43414 & 5852 & 25138  & -457042  
    \end{tabular}
    \caption{Allemagne, $2h$ battery, $\Delta =2$ : backtest gain on a single battery in 2023}
    \label{tab:al2023Idec2}
\end{table}
\end{comment}
\begin{table}[H]
    \centering
    \begin{tabular}{rr|rr|rr|rr|rr}
    \toprule
    Year &  $\hat{N}$ & \multicolumn{2}{c}{1} & \multicolumn{2}{c}{20} & \multicolumn{2}{c}{50} & \multicolumn{2}{c}{100} \\
    & Model & Depth & N.D. & Depth &N.D. & Depth & N.D. & Depth & N.D. \\
    \midrule
    \multirow{2}{*}{2022} & Det.
    & 86591 & 86605 & 140591 & 80845 & 79625 & 56381 & -18218 & -101280 \\
    & Sto. 
    & 106882 & 106813 & 101778 & 101284 & 90521 & 79625 & 68123 & -56460 \\
    \midrule
    \multirow{2}{*}{2023} & Det. 
    & 44812 & 44416 & 36246 & 33833 & 8164 & -35606 & -73747 & -584819 \\
    & Sto.
    & 61823 & 61824 & 55285 & 53964 & 43414 & 5852 & 25138 & -457042 \\
    \bottomrule
    \end{tabular}
    \caption{Backtest gain (euros per battery) for the optimisation of $\hat{N}$ $2$h batteries for the years 2022 and 2023 on the German market for the different models. The market position is taken $2$ hour before maturity. Det. is for Determinist, Sto. for Stochastic, and N.D. for No depth.}
    \label{tab:al2022_2023_combined}
\end{table}

\end{document}